\documentclass[12pt]{article}
\usepackage{graphicx}

\begin{document}
\begin{titlepage}
%\baselineskip=3.1truemm
%\columnsep=.5truecm
%
%\newenvironment{lefteqnarray}{\arraycolsep=0pt\begin{eqnarray}}
%{\end{eqnarray}\protect\aftergroup\ignorespaces}
%\newenvironment{lefteqnarray*}{\arraycolsep=0pt\begin{eqnarray*}}
%{\end{eqnarray*}\protect\aftergroup\ignorespaces}
%\newenvironment{leftsubeqnarray}{\arraycolsep=0pt\begin{subeqnarray}}
%{\end{subeqnarray}\protect\aftergroup\ignorespaces}
%

% Running titles

%\markboth{\eightrm THE CHEMICAL POTENTIAL OF A LENNARD JONES FLUID}
%{\eightrm V. {\v C}ELEBONOVI{\' C}}

%{\ }

%\publ

%\type

{\ }

% Title
\begin{center}
\centerline{A condensed matter analogy of impact crater formation}

%\footnote{Published in Serbian Astron.Journal,{\bf181},pp.51-55,(2010)}
\end{center}
% Authors
\begin{center}
\centerline{V.Celebonovic}			
\end{center}
\vskip5mm

% Address
\begin{center}
\centerline{Inst. of Physics,Univ. of Belgrade,Pregrevica 118,11080 Zemun-Belgrade,Serbia}
\vskip3mm 
\centerline{vladan@ipb.ac.rs}
\end{center}

% Received and Accepted dates

 %Abstract  

\begin{abstract}
Impact craters exist on various solid objects in the planetary system. A simplified analogy of the process of their formation is here analyzed by standard solid state physics and the so called dynamic quantized fracture mechanics. An expression which links the crater volume to the parameters of the impactor and the target is obtained within the two approaches. For low impactor energy, this expression is of the same mathematical form as the one  resulting from recent experiments.It is shown that the formation of an impact crater is possible even without heating of the target, if the critical stress in the target satisfies certain conditions. The critical value of the stress needed for the occurence of a fracture is calculated for three craters: two terrestrial and one lunar crater. The approach presented here uses only measurable material parameters, and is therefore more realistic than the treatement of the same problem using the cohesive energy of materials.

\end{abstract}
%\footnote{Published in Serbian Astron.Journal,{\bf181},pp.51-55,(2010)}
\end{titlepage}
% Keywords (see keywords.pdf file)

%\keywords{Physical data and processes--Equation of state}

%\begin{multicols}{2}

% Sections

\section{Introduction}

The existence of some craters on the surface of the Earth is due to impacts of small bodies in the planet. A list of such craters is avaliable at the web site http://www.passc.net/EarthImpactDatabase/index.html. The study of these craters, and the constraints which their existence places on the impactors has become a separate field of research in planetary science (for example [1]). Two probably best known examples of such events are the impacts which have led to the formation of the Barringer crater in Arizona, and the Tunguska event of 1908. 

What can be concluded about the impactors by combining astronomical data with results of solid state physics? It was recently shown ( [2] and related work) that by using basic principles of condensed matter physics, it becomes possible to derive an expression for the product $\rho_{1} r_{1}^{3} v_{1}^{2}$, where the three symbols denote the mass density, radius and speed of the impactor. In that calculation, the notion of cohesion energy of a solid was used. This quantity is defined as the energy needed to transform a sample of a solid into a gas of widely separated atoms. It is not easy to measure experimentally, and it is not related to the practical strength of solids, which is regulated by their resistance to fracture [3]. 
It this letter we investigate a simplified  analogy of the formation of an impact crater. The object we investigate is a hole of given dimensions, resulting from the impact of an external object in a material with known parameters. Using the dimensions of the "hole" as the final result of the impact and parameters of the target, what can be concluded about the impactor?   

The aim of this paper is to derive an expression for $\rho_{1} r_{1}^{3} v_{1}^{2}$ by using the notion of stress instead of the notion of cohesive energy. The stress is defined as the ratio of the force applied on a body to the cross section of the surface of a body normal to the direction of the force [4]. In order for a crater to form in the target as a result of an impact, stress must become sufficiently high so as to allow the formation of a fracture in the material of the target. The following section contains a brief reminder of the required definitions and previous results, while the third part is devoted to the calculations. A discussion of the results and the conclusions are presented in the fourth and fifth sections. 

The approach used in the present work is based on general principles of solid state physics, and it can be applied to any material. The main tool used in derivation of scaling theories, discussed for example in [6],is dimensional analysis. See also [5]. According to the scaling theory, one of the factors on which the volume of an impact crater depends is the gravitational acceleration at the surface of the target. In the section devoted to discussion it will be shown that the calculation reported here leads to the same result. 

This paper uses solid state physics throughout. This means that it is assumed that the material of the target does not melt in the impacts,which implies small kinetic energies of the impactors. Various aspects of heating in mutual collisions of solids has recently been discussed ; examples are [7], [8].

Apart from fundamental interest, studies of impact craters and the projectiles which  have made them have a very "practical" motivation. Impacts into the Earth  have been occuring throughout the history of our planet,and will occur again. An impact, if sufficiently energetic, could have serious consequences for the region where it occurs, or the planet as a whole, so predicting the outcome of such events is highly important. At the time of this writing, the last example of such an event is the impact of a small asteroid designated $2014AA$ into the Atlantic on January 2,2014. 

\section{The basics}
The basic physical condition for the formation of an impact crater is that a fracture must form in the material of the target as a result of the impact. The material parameter used in evaluating the possibility of the occurence of a fracture is the value of the stress existing on a crack preexisting in a material. This critical value of the stress can be evaluated in two ways: by standard solid state physics [9] and within the so called dynamic quantized fracture mechanics (DQFM) [10], [11].

  It can be shown in standard solid state physics (for instance [9]) that the critical value of the stress needed for the occurence of a fracture in a material is
  \begin{equation}
  \label{eq.1}
  \sigma_{C} =  \frac{1}{2} [\frac{E\chi\tau}{r_{0} w}]^{1/2}
  \end{equation}
where $E$ is Young's modulus of the material,$\chi$ is the surface energy,$\tau$ the radius of curvature of the crack,$r_{0}$ the interatomic spacing at which the stress becomes zero and $w$ denotes the length of the crack. A simple analysis shows that $\sigma_{C}$ has the dimensions of pressure,which means that the stress multiplied by a volume has the dimensions of energy. 

An alternative possibility of estimating the stress needed for fracturing a material is given by the dynamic quantized fracture mechanics (DQFM for short). The difference between $DQFM$ and the usual approach used in material science is that $DQFM$ introduces geometry in studies of scaling laws in material science (for example [11]). Considering that the occurence of a fracture in a material is a sign of its failure, it can be shown in $DQFM$ that the stress necessary for the occurence of a failure is given by [11]:
\begin{equation}
\label{eq.2}
	\sigma_{f} = K_{Ic}[\frac{1+(\frac{\rho_{0}}{2q})}{\pi (l_{0}+(q/2))}]^{1/2}
\end{equation}
where $K_{Ic}$ denotes the fracture toughness, $\rho_{0}$ is the radius of curvature of the tip of the crack of length $l_{0}$ and $q$ is the length of the so called fracture quantum. 
The kinetic energy of the impactor in the moment of impact is used for fracturing and heating the material of the target. This means that:
\begin{equation}
\label{eq.3}
E_{k}=\sigma_{c} V + C_{V} V (T_{1}-T_{0})
\end{equation}
 where $V$ is the volume of the crater formed as a consequence of the impact, $C_{V}$ is the heat capacity of the target material, and $T_{0}$ the intial temperature of the target. The volume of the crater depends on its form. In line with recent experimental results, such as [12], it was taken that the crater volume is given by
%It can be assumed that it has the form of a half of a rotational ellipsoid, which means that the volume is
\begin{equation}
\label{eq.4}
V=\frac{1}{3}\pi b^{2} c
\end{equation}
where $b$ denotes the radius of the "opening" of the crater and $c$ its depth. 

The specific heat of a solid is given by
\begin{equation}
\label{eq.5}
	C_{V} \cong 3 N\nu k_{B} [1-(\frac{1}{20})(\frac{T_{D}}{T})^{2}+ (\frac{1}{560})(\frac{T_{D}}{T})^4]
\end{equation}
where $T_{D}$ is the Debye temperature,$\nu$ is the number of particles in the elementary crystal cell,$N$ is the number of elementary crystal cells in the specimen,and $k_{B}$ is Boltzmann's constant.

\section{Calculations} 
\subsection{The standard treatment} 
It will be assumed that the impactor is a sphere of radius $r_{1}$ made up of a material of density $\rho_{1}$, having velocity $v_{1}$ in the moment of impact. This means that its kinetic energy is given by $E_{k}=\frac{2\pi}{3}\rho_{1} r_{1}^{3} v_{1}^{2}$.

It follows from eq.(3) that in a collision the material of the target heats to a temperature $T_{1}$ given by
\begin{equation}
\label{eq.6}
T_{1} = T_{0} + \frac{1}{C_{V}} (\frac{E_{k}}{V}- \sigma_{C})
\end{equation}
It follows from eq.(6) that $V = E_{k}/((T_{1}-T_{0}) C_{V}+\sigma_{C})$. Using eq.(4), it follows that the ratio $E_{k}/V$ is given by 
\begin{equation}
\label{eq.7}
\frac{E_{k}}{V} = \frac{2}{b^{2} c} \rho_{1} r_{1}^{3} v_{1}^{2}
\end{equation}
which leads to
\begin{equation}
\label{eq.8}
T_{1}-T_{0} = \frac{1}{C_{V}}\times(\frac{2}{ b^{2} c} \rho_{1} r_{1}^{3} v_{1}^{2} - \sigma_{C})
\end{equation} 
As the target heats up as a consequence of the impact,$T_{1}>T_{0}$, which implies that $\rho_{1} r_{1}^{3} v_{1}^{2} > (3 V/2\pi) \sigma_{C}$. Expressing $T_{1}- T_{0}$ as $\alpha T_{0}$ with $\alpha\geq 0$ and using eq.(4), it follows that 
\begin{equation}
\label{eq.9}
\rho_{1} r_{1}^{3} v_{1}^{2} = \frac{3 V}{ 2 \pi} \times(\alpha T_{0} C_{V}+\sigma_{C})
\end{equation}
and finally
\begin{equation}
\label{eq.10}
V=\frac{2 \pi}{3}\frac{\rho_{1} r_{1}^{3} v_{1}^{2}}{\alpha C_{V} T_{0} + \sigma_{C}}
\end{equation}
\subsection{The DQFM treatement} 
The DQFM theory introduces geometry into considerations of the occurence of fracture and failure of solids. Taking the ratios $\rho_{0}/(2q)$ and $q/(2 l_{0})$ in eq.(2) as small parameters, then developing up to first order terms, it follows that
\begin{eqnarray}
\label{eq.11}
\sigma_{f}\cong\frac{K_{Ic}}{\sqrt{\pi l_{0}}}\times[1+\frac{1}{2}\times(\frac{\rho_{0}}{2 q}-\frac{q}{2 l_{0}})]
\end{eqnarray}
The energy ballance has the same form as in the preceeding case
\begin{equation}
\label{eq.12}
E_{k} = \sigma_{f} V + C_{V} V (T_{1}-T_{0})
\end{equation}
which gives
\begin{equation}
\label{eq.13}
V=\frac{2 \pi}{3}\frac{\rho_{1} r_{1}^{3} v_{1}^{2}}{\alpha C_{V} T_{0} + \sigma_{f}}
%\rho_{1} r_{1}^{3} v_{1}^{2} = \frac{3 V}{2 \pi}\times(\alpha T_{0} C_{V} + \sigma_{f})
\end{equation}
Solving eq.(13) for $V$ and then inserting eq.(11) gives the following approximate relationship between the parameters of the impactor $(\rho_{1},r_{1},v_{1})$ and those of the material of the target $(\sigma_{f},\rho_{0},l_{0},q)$. 
\begin{eqnarray}
\label{eq.14}
V \cong\frac{2 \pi}{3} \frac{\rho_{1} r_{1}^{3} v_{1}^{2}}{\alpha T_{0} C_{V}+\frac{K_{Ic}}{\sqrt{\pi l_{0}}}\times[1+\frac{1}{2}\times(\frac{\rho_{0}}{2 q}-\frac{q}{2 l_{0}})]}
\end{eqnarray}
\section{Discussion} 
\subsection{Comparison with laboratory data}
Equation (10) can be reexpressed as a linear function
\begin{equation}
\label{eq.15}
V = E_{k}/(\alpha C_{V}T_{0}+\sigma_{C})
\end{equation}
which means that the crater volume is a linear function of the kinetic energy of the impactor. On the other hand, raw experimental data on crater volumes and the impactor energies in [12], can be fitted by an equation of the form $V[m^{3}] = V_{0}\times Exp[E[J]/c]$, with $V_{0} = (4\pm2)\times 10^{- 7} m^{3}$ and $c = (583\pm56) J$. For sufficiently low energies $E$, this exponential expression reduces to the form $V-V_{0}\cong(V_{0}/c)E$. Combining with results of the calculations reported here, it follows that $V_{0}/c = 1/(\alpha C_{V}T_{0}+\sigma_{C})$ The implication is that the results of the calculations reported here are relevant to low kinetic energies of the impactors. 
\subsection{Standard treatment} 
Equations (10) and (13) represent the relationships between the parameters of the impactor and those of the target, obtained within two slightly differing theoretical frameworks: standard solid state theory and the so called $DQFM$ theory. Although their mathematical form is identical, eq.(13) is physically more complex, as it takes into account the geometric parameters of the target material by eq.(2).  

Several physically interesting limiting cases are visible from eq.(10). In the case $\alpha\rightarrow 0$, it follows that
\begin{equation}
\label{eq.16}
\lim_{\alpha\rightarrow 0}V = \frac{2 \pi}{3 \sigma_{C}} \rho_{1} r_{1}^{3} v_{1}^{2}
\end{equation}
which corresponds to an impact without the heating of the target,
and, if $\sigma_{C}\rightarrow 0$, which means that the stress needed for the occurence of a fracture is very small, 
\begin{equation}
\label{eq.17}
\lim_{\sigma_{C}\rightarrow 0}V = \frac{2 \pi}{3} \frac{\rho_{1} r_{1}^{3} v_{1}^{2}}{\alpha T_{0} C_{V}}
\end{equation} 
Finally, if the stress needed for the occurence of a fracture in the target material is high, it follows that
\begin{equation}
\label{eq.18}
\lim_{\sigma_{C}\rightarrow \infty} V = 0
\end{equation} 
which  means that in the case of very high values of $\sigma_{C}$ the volume of an impact crater tends to zero.  
\subsection{Treatment within DQFM} 
In the $DQFM$ theory the final expression for the volume of a crater formed as a result of an impact is similar to eq.(13)
which is, to first order, approximated by eq.(14). 

If the fracture toughness is extremely high, $\sigma_{f}$ is high, and it follows that $\lim_{K_{Ic}\rightarrow \infty}V=0$. This is physically expectable,as $K_{Ic}$ characterizes the ability of a material with a crack to resist fracture. In the case when both the fracture toughness and the heating in the impact are small, $\lim_{K_{I_{c}}\rightarrow 0,\alpha \rightarrow 0}V=\infty$
%\begin{equation}
%\label{eq.20}
%\lim_{K_{I_{c}}\rightarrow 0,\alpha \rightarrow 0}V = \infty
%\end{equation} 
the volume of the crater formed can be arbitrarily large. If the length of the fracture quantum is arbitrarily small,
the volume of the impact crater tends to zero. Finally, if $l_{0}\rightarrow\infty$, the volume of the crater can get abitrarily large.
\subsection{Comparison with scaling theory} 

Standard scaling theory shows,for example [6],that the volume of a crater formed as a consequence of an impact depends on the mass density of the target material. The same conclusion is reached in the present paper, but from results of condensed matter physics. This dependence is implicite in eqs.(10) and (13). 
Young's modulus $E$ is given by [3] 
\begin{equation}
\label{eq.19}
E = \mu \frac{3 \lambda + 2\mu}{\lambda + \mu}
\end{equation} 
where $\lambda$ and $\mu$ are Lame's paremeters. Parameter $\lambda$ is related to the velocities of pressure and shear waves in a material by
\begin{equation}
\label{eq.20}
\lambda = \rho\times(v_{P}^{2}-2 v_{S}^{2})
\end{equation}
The velocities are given by $v_{S}^{2}\rho = \mu$ and $v_{P}^{2}\rho = \lambda+2\mu=B_{0}+(4/3)\mu$, where $B_{0} = \rho \partial P/\partial\rho$ is the bulk modulus. They are related to the pressure $P$, density $\rho$ and bulk modulus $B_{0}$ of a material by [13]
\begin{equation}
\label{eq.21}
v_{P}^{2} = \frac{9 B_{0}}{5\rho}\times [1-\frac{8 P}{9 B_{0}}]
\end{equation}
and 
\begin{equation}
\label{eq.22}
v_{S}^{2} = \frac{3}{5\rho}\times [B_{0}-2 P]
\end{equation}
which leads to
\begin{equation}
\label{eq.23}
\lambda = \frac{3 B_{0}}{5}\times [1+\frac{4 P}{3 B_{0}}]
\end{equation}
and 
\begin{equation}
\label{eq.24}
\mu = v_{S}^{2}\times\rho = \frac{3}{5}\times(B_{0}-2 P)
\end{equation}

Finally one gets
\begin{equation}
\label{eq.25}
E=3 \mu \frac{1}{1+\frac{\mu}{3 B_{0}}} = \frac{9}{2} B_{0}\times\frac{B_{0}-2P}{3B_{0} - P}
\end{equation}
as the result for Young's modulus of a material under pressure $P$ and having bulk modulus $B_{0}$. Inserting into eq.(1), it follows that 
\begin{equation}
\label{eq.26}
\sigma_{C}=\frac{3}{2^{3/2}}(\frac{\chi\tau B_{0}}{r_{0} w}\times\frac{B_{0}-2 P}{3B_{0}-P})^{1/2}
\end{equation}

The pressure is related to the density by the equation of state, so eq.(25) gives in fact the density dependence of Young's modulus. %The exact analytical form of eq.(25) depends on the form of the equation of state of the material. 
If, for example, one assumes the applicability of the Birch-Murnaghan equation of state, the pressure is approximately given by:
\begin{equation}
\label{eq.27}
P(\rho)\cong\frac{9 B_{0} B_{0}'}{8}(\frac{\rho}{\rho_{0}})^{3} (1-\frac{16}{3 B_{0}'}-...)
\end{equation}
where $B_{0}' = \partial B_{0}/\partial \rho$. Combined with eqs (25) and (10), this means that the volume of a crater formed in an impact is also a function of the density of the target. 

As the mean density $\rho$ of an object of radius $R$ is proportional to the gravitational acceleration $g$ at the surface by: $\rho = 3g/(4\pi\gamma R)$, where $\gamma$ is the gravitational constant, this implies that the volume of an impact crater depends on the gravitational acceleration on the suface of the target, which is in line with results of the scaling theory. 

Equations (10) and (13) can be used in two ways: if the volume of the crater is known, and the parameters $r_{1}$, $v_{1}$ and $\rho_{1}$ can somehow be estimated, the value of $\sigma_{C}$ can be calculated from these equations assuming various values of $\alpha$. They can also be used to study the dependence of the volume of a crater on any of the parameters contained in these expressions.

Equations (13) and (15) express the volume of the crater formed in an impact as a function of the parameters of the impactor and of the material of the target.
With all the other parameters fixed, the  volume of the crater formed in the target is inversely proportional to the critical value of the stress needed for the formation of a fracture in the material of the target and the heat capacity of the material of the target. A further implication is that the volume of such a crater is proportional to the length of a crack preexisting in a material, and inversely proportional to Young's modulus of the target.

The fact that a crater can be formed in an impact without heating of the target may seem strange at first. However, this is physically plausible, assuming that the heat capacity of the material of the target is high enough and, at the same time, the impact is slow enough, so that the kinetic energy of the impactor can not heat the target, but is spent on fracturing the materal of the target. Photographs of the lunar surface, taken from various space probes and telescopes on the Earth illustrate  this conclusion, as there are many small craters which show no sign of heating and/or melting in their formation.Some of these photographs can be accessed at the address:
http://www.lpi.usra.edu/lunar/.

\section{Test examples}

Expressions derived in the previous section will here be applied to three "experimental" examples: two terrestrial and one lunar crater, with the aim of testing the applicability of the expressions derived in this paper. It will be taken that the initial temperature of the target is $T_{0} = 300 K$. 

As a first example take the Barringer crater in Arizona. For an extensive review of this object see [14]. It has been estimated that the total volume of material ejected in this impact is $V \cong 7.6\times 10^{7} m^{3}$ [15]. The most abundant mineral on this site is $SiO_{2}$ [14], for which $\sigma_{C}\cong 8.4\times10^{9} Pa$ [16]. Results obtained in [2] indicate that $\rho_{1}r_{1}^{3}v_{1}^{2}\cong 4\times10^{17} J$. Solving eq.(9), with $C_{V}\approx 2 \times 10^{6} J/m^{3}K$, then gives $\alpha\approx4.37$, which means that the material at the impact site heated up to a temperature $T_{1}=\alpha\times T_{0}\approx1300 K$. The melting temperature of $SiO_{2}$ is approximately $T_{m}\approx1900 K$. As $T_{1}<T_{m}$ solid state physics is applicable.

The shape of the function $V(\alpha)$ for this particular case is shown on fig.1. The volume on figure 1 is normalized to the volume of the Barringer crater. Clearly, the higher $\alpha$ the smaller the volume of the crater resulting from the impact becomes.
For comparison purposes, the same calculation was attempted with data discussed in [17]. Taking these data at "face value", the value of $V$ from [17], and the value of $\sigma_{C}$ from the previous paragraph, it follows that $\rho_{1}r_{1}^{3} v_{1}^{2} = 9.216\times 10^{15}J$, while $3V\sigma_{C}/2 \pi= 3.0481\times 10^{17}$. This means that the condition $\rho_{1} r_{1}^{3} v_{1}^{2} > (3 V/2\pi) \sigma_{C}$ is not fulfilled. However, increasing the values of $\rho_{1},r_{1},v_{1}$ by,respectively, $5$ and $30$ percent compared to values discussed in [17], and assuming that $\alpha = 0.2$ , one gets $\sigma_{C}\cong 1.106\times 10^9 Pa$. The difference between this value and the one obtained earlier in this work is obvious, but can be traced to the different values of $\rho_{1},r_{1},v_{1}$ used in the two calculations.
\begin{center}
\begin{figure}
\includegraphics[width=9cm]{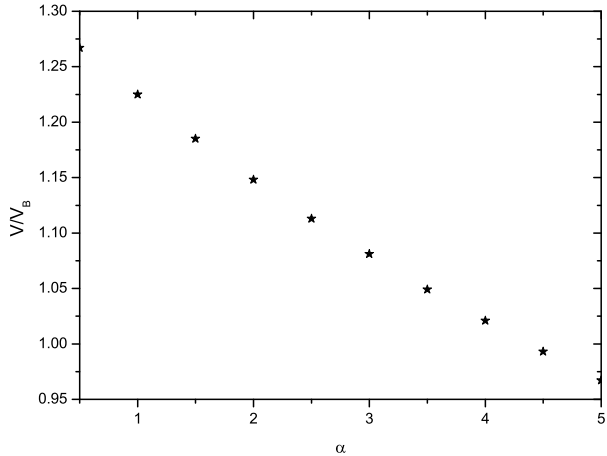}
\caption{\it{The volume of an impact crater as a function of $\alpha$, normalized to the volume of the Barringer crater}}
\end{figure}
\end{center}
Another terrestrial example is the Kamil Crater, on the border of Egypt with Sudan. 
Its age is estimated at around 5000 years before present, which makes it remarkably fresh by geological criteria [18]. The dimensions of the crater have been measured, and its volume can be calculated by eq.(4) as $V\cong 8863.5 m^{3}$ . Using data given in [18], gives $\rho_{1}r_{1}^{3} v_{1}^{2} = 9.11\times 10^{11} J$, which implies that $\sigma_{C}+\alpha\times C_{V}\times T_{0}=2.153\times10^8$. Assuming that $\alpha=0.2$,and that the material of the target is quartz sand for which $C_{V}\cong 9.96 \times 10^{5} J/m^{3} K$, one gets $\sigma_{C}\cong 1.56\times 10^{8} J/m^{3}$ .   

An interesting example concerns impact craters on the surface of the Moon. A multitude of data on them has been accumulated in the Apollo program (for example [19]). Paper [19] contains detailed results on three impacts recorded during the functioning of the seismological stations left on the lunar surface. One of these impacts was recorded on January 25,1976. Using data given in [19] for this particular impact, one gets that $\rho_{1} r_{1}^{3} v_{1}^{2} = 4.05\times 10^{12} J$, and the volume can be calculated from the measured dimensions as $V=2.18\times 10^{4}m^{3}$. Using $C_{V}=2.76\times10^{6}J/m^{3} K$ for the lunar material from [20] and assuming thet $\alpha = 0.2$,it follows that $\sigma_{C} = 2.23\times 10^{8} J/m^{3}$. This value is an order of magnitude lower that the result for the Barrringer crater. Physically, it means that the critical value of the stress needed for the occurence of a fracture in the lunar surface is lower than the corresponding value for the surface of the Earth. Qualitatively speaking, this is in agreement with the known fact that the lunar surface is "softer" than the surface of the Earth (for example [21]).  

\section{Conclusions}
In this letter we have discussed to some extent the process analogous to the formation of impact craters on the surfaces of the solid object in the planetary system. This problem is of high practical importance, because impacts of small bodies into the Earth have been occuring and will occur again,as testified by the impact of asteroid $2014AA$ in the Atlantic on January 2, 2014. 

The approach to the problem discussed here, differs from the one used in [2] in the physics used. In [2] the calculation was performed using the cohesion energy (of the material of the target). The treatement used here is a distinct advantage because it uses only measurable quantities-various parameters of the target and the impactor. The values of the critical stress needed for the occurence of a fracture were calculated for three impact craters: two terrestrial and one lunar,and the values obtained differ by one order of magnitude. This difference can be ascribed to two factors: varying quality of initial data, and the real physical difference of the materials on the three crater sites. Note that the value of $\sigma_{C}$ of a material depends on the chemical composition. In the calculation for the Barringer crater it was assumed that the most abundant mineral there is $SiO_{2}$. Changing this assumption would change the values of $\sigma_{C}$ and $C_{V}$.  It will be attempted to improve the results of the approach discussed in this letter by taking into account more physical details of the process of formation of impact craters. Note that within the $DQFM$ theory the value of the critical stress needed for a material to fracture, and therefore a crater to form, depends  also on the geometrical parameters of cracks existing in the material. This conclusion potentially opens the possibility of applying the $DQFM$ in terrestrial laboratory work. Details will be discussed in the future.

\section{Acknowledgement} 

The preparation of this work was financed by the Ministry of Education, 
Science and Technological Developement of Serbia under its project 141007. 

%\begin{thebibliography}{00}
{}
%\end{thebibliography}

\begin{thebibliography}{00}

\bibitem{[1]} 

Melosh,H.J. 1989, Impact Cratering,A Geologic Process, Oxford:

Clarendon Press 

%\bibitem[Melosh{1989}]{mel89} Melosh,H.J. 1989, Impact Cratering,A Geologic Process, Oxford: Clarendon Press 

\bibitem{[2]} 

Celebonovic,V. 2013, Rev.Mex.Astron.Astrophys,49,221


\bibitem{[3]}  Marder,M.P. 2010, Condensed Matter Physics, Hoboken,NJ: John Wiley and Sons, Inc.,295

\bibitem{[4]} Celebonovic,V. 2007, in Solar and Stellar Physics Through

 Eclipses, ed by O.Demircan,S.O.Selam and B.Albayrak,

 San Francisco: Astron.Soc.Pacific,20 


\bibitem{[5]} Holsapple,K.A. \& Housen,K.R. 2007, Icarus, 187, 345

\bibitem{[6]} Holsapple,K.A. 1993, in Ann.Rev.Earth and Planet Sci.,21,333

\bibitem{[7]} Davison,T.M., Collins,G.S. \& Ciesla,F.J. 2010, Icarus,208, 468


\bibitem{[8]} Celebonovic,V. 2012, Serb.Astron.J.,184, 83 

 
\bibitem{[9]} Tiley,R. 2004, Understanding Solids: The Science of Materials, Hoboken,NJ: John Wiley and Sons, Inc., 548

\bibitem{[10]} Carpinteri,A.\& Pugno,N. 2005, Nature Materials, 4, 421.

\bibitem{[11]} Pugno,N. 2006, Int.J. Fract, 140, 159

\bibitem{[12]} Suzuki,A., Hakura,S., Hamura,T \& et al 2012, J.Geophys.Res., 117, E08012 


\bibitem{[13]} Gaurav,S.,Subramanian,S.S.,Singh,S.P \& Sharma,B.P.,
 
2012, arxiv 1207:0283


\bibitem{[14]} Kring,D., 2007, Guidebook to the Geology of Barringer Meteorite Crater,Arizona, Houston: LPI, Contribution 1355

\bibitem{[15]} Roddy,D.J., Boyce,J.M., Colton,G.W.\& Dial A.L.,Jr. 1975, Proc 6th Lunar Sci.Conf., 2621

\bibitem{[16]} Petersen,K.E. 1982, Proc.IEEE, 70, 420

\bibitem{[17]} Melosh,H.J. and Collins,G.S. 2005, Nature,434, 157

\bibitem{[18]} Folco L. Di Martino M, El Barkooky A,et al.,2010,Science, 329, 804

\bibitem{[19]} Gudkova T.V.,Lognonne Ph.and Gagnepain-Beyneix J.,2011, Icarus, 211,1049

\bibitem{[20]} Hemingway B.S.,Robie R.A. \& Wilson W.H. 1973, Proc. Lun.Sci. Conf.,4,2481


\bibitem{[21]} Mitchell,J.K., Houston,W.N.,Carrier,W.D.III \& Costes,N.C. 1974, 

Apollo soil mechanics experiment S-200, NASA: CR 134306

%\references

\end{thebibliography}
\end{document}